\documentclass{ws-acs}

\begin{document}

\markboth{S.~Bouzat \& D.~H.~Zanette} {Survival and extinction in
the replicator model:  Dynamics and statistics}

\catchline

\title{Survival and extinction in the replicator model:
Dynamics and statistics}

\author{SEBASTI\'AN BOUZAT and DAMI\'AN H. ZANETTE}

\address{Consejo Nacional de Investigaciones Cient\'{\i}ficas y
T\'ecnicas\\ Centro At\'omico Bariloche and Instituto Balseiro \\
8400 San Carlos de Bariloche, R\'{\i}o Negro, Argentina
}

\maketitle

\pub{Received (Day Month Year)}{Revised (Day Month Year)}
{Accepted (Day Month Year)}

\begin{abstract}
We study the multi-species replicator model with linear fitness
and random fitness matrices of various classes. By means of
numerical resolution of the replicator equations, we determine
the survival probability of a species in terms of its average
interaction with the rest of the system. The role of the
interaction pattern of the ecosystem in defining survival and
extinction probabilities is emphasized.

\end{abstract}

\section{Introduction}

Consider an ecosystem formed by $N$ biological species, with
populations $n_1,\dots,n_N$. Over time scales where the effect of
mutations can be neglected, the population of species $i$ is
governed by the evolution equation
\begin{equation} \label{1}
\dot n_i =f_i n_i,
\end{equation}
where $f_i$ is the (Fisherian) fitness of that species
\cite{Fisher,Michod}. Equation (\ref{1}) assumes that evolution
is continuous in time and that populations are large enough as to
neglect the effect of random drift \cite{Nagylaki}. The
interaction between species in the ecosystem is described by the
fitnesses $f_i$, which generally depend on the whole set of
populations. They may also include the effect of migration from
and toward the system. It can be straightforwardly shown that the
frequency of species $i$ in the whole population, $x_i=n_i /
\sum_j n_j$, satisfies the evolution equation
\begin{equation}
\label{repf}
\dot x_i=x_i \left( f_i- \sum_j x_j f_j \right) .
\end{equation}
Assuming that the fitnesses can be given as functions of
$x_1,\dots,x_N$ gives a closed set of equations for the
frequencies, usually referred to as the {\it replicator
equations} \cite{hof}.

Note that the replicator equations admit an alternative
interpretation as the evolution equations for the populations
--instead of the frequencies-- of an $N$-species ecosystem. Since
for all times $\sum_i x_i=1$ (see also Sect. \ref{ii}), this
would correspond to an ecosystem with fixed total population,
presumably saturating the carrying capacity of its environment.
Within this interpretation, thus, the replicator equations
describe an ensemble of biological species with global
competition, which limits the growth of the whole population to
the carrying capacity of the environment, and subject to
interactions  that affect the individual fitnesses.

The replicator model has found many applications in biological
evolutionary problems at various levels. It has been first
considered in connection with the evolution of biomolecular
concentrations in the early stages of life on the Earth
\cite{Eigen}. In the frame of game theory, in close connection
with biological evolution \cite{Maynard}, it has been used to
describe the dynamics of strategy frequencies in multi-player
games \cite{Taylor}. As a basic model of an ecosystem, it was
discussed by Schuster and Sigmund \cite{Schuster}, who coined the
term {\it replicator}. The discrete-time version of Eqs.
(\ref{repf}) \cite{Samuelson} constitutes a well-known model for
the evolution of genetic frequencies in asexual haploid
populations \cite{Nagylaki}. Most of these applications assume
that the fitnesses depend linearly on the frequencies, $f_i
=\sum_j a_{ij} x_j$, so that the replicator equations reduce to
\begin{equation}
\label{sys}
\dot x_i=x_i \left(\sum_j a_{ij} x_j - \sum_{j,k} a_{jk}x_jx_k
\right).
\end{equation}
The matrix $A = \{ a_{ij} \}$ is frequently called {\it fitness
matrix} and, from the viewpoint of game theory, plays the role of
a payoff matrix. It can be shown that the replicator equations
with linear fitness are formally equivalent to a multi-species
Lotka-Volterra model \cite{hof}.

In spite of the large potential of the replicator equations to
model evolutionary biological systems, the dynamical properties
of their solutions are poorly understood. Besides the specific
applications quoted above, generic mathematical properties of the
replicator model are known for very special situations only
\cite{hof,Weibull}. Typically, they involve a small number of
biological species with quite unrealistic interaction patterns.
Real ecosystems, on the other hand, usually involve tens of
species with hundreds of food-web links \cite{eco1,eco2}. Our aim
in this paper is to approach this complex reality by exploring the
replicator dynamics for large ecosystems with nontrivial
interaction patterns. We deal with the problem from a statistical
point of view and perform extensive numerical calculations,
focusing on a probabilistic description of survival and
extinction of species in terms of the parameters that define
their interaction.

In the next two sections we present --mainly for completeness and
subsequent reference-- various, rather general mathematical
properties of the replicator equations, including symmetry
properties and stability conditions for special classes of random
interaction patterns. Then, in the main part of the paper, we
numerically study more complex situations for the case of linear
fitness, including multi-diagonal and disordered fitness
matrices. Our results, which we interpret in connection with the
average interaction of each species with the rest of the
ecosystem, are summarized and discussed in the last section.

\section{Invariance properties and symmetries} \label{ii}

For future reference, let us begin discussing the invariance
properties of the replicator model (\ref{repf}). In the first
place, since the variables $x_i$ are defined as population
fractions, $x_i=n_i/\sum_jn_j$, the conditions $x_i \ge 0$ and
$\sum_i x_i=1$ must hold at all times.  Equations (\ref{repf}) are
compatible with these conditions, because the set $S_N=\{ {\bf x}=
(x_1,...,x_N) / x_i\ge 0\ \forall i \mbox{ and }\sum_i x_i=1\} $
is invariant under their action. In fact, the transverse
component of $\dot {\bf x}$ vanishes on the boundary of $S_N$, and
\begin{equation}
\label{uno}
\sum_i \dot x_i =\left( 1 - \sum_i x_i\right) \sum_j x_j f_j
\end{equation}
identically vanishes in $S_N$. Consequently, an initial condition
in $S_N$ generates a trajectory which remains within this set.
From now on, we restrict the analysis of Eqs.~(\ref{repf}) to
$S_N$, where the dynamics is effectively  $N-1$-dimensional. In
fact, $S_N$ is an $N-1$-dimensional polyhedron embedded in the
$N$-dimensional Euclidean space.

Equations (\ref{repf}) have moreover two important symmetries.
First, multiplication of all the fitnesses  by the same constant
$f_i\to c_1f_i$, is equivalent  to a linear change of the
temporal scale, $t \to c_1 t$. Second, addition of  the same
constant to all the fitnesses, $f_i\to f_i +c_2$ leaves the
equations invariant. This second symmetry implies that any
meaningful result regarding the replicator dynamics will be given
in terms of differences between fitnesses, rather than in terms of
their individual values. Note, in fact, that Eqs.~(\ref{repf}) can
be rewritten as
\begin{equation}
\dot x_i = x_i \sum_j (f_i-f_j)x_j.
\end{equation}
For linear fitness, Eqs.~(\ref{sys}), the above symmetries
respectively amount to multiplying and adding a constant to each
element of the fitness matrix $A$, $a_{ij}\to c_1 a_{ij}$ and
$a_{ij} \to a_{ij}+c_2$.

\section{Linear fitness: Ecological attitude and behavior} \label{iii}

In the following, we focus the analysis on the case of linear
fitness, $f_i=\sum_j a_{ij} x_j$. For the sake of simplicity, the
restriction of Eqs.~(\ref{sys}) to $S_N$ will be referred to as
the replicator equations.

The  two summations in the right hand side of Eqs.~(\ref{sys})
couple the dynamics of the species in different ways. The first
summation, $\sum_j a_{ij} x_j$, coincides with the fitness of
species $i$, and is a direct measure of the effect of the whole
ecosystem on the reproductive success of that species. The
coefficient $a_{ij}$, in fact, weights the contribution of
species $j$ to the fitness of $i$. The second summation,
$\sum_{kl} a_{kl}x_k x_l$ is the average fitness in the
ecosystem, and acts as a global coupling. This ``interaction''
--a negative contribution for positive $a_{kl}$-- is responsible
for the preservation of the normalization of $\sum_ix_i$. The
invariance of the system under summation of a constant to the
fitness, quoted in Sect. \ref{ii}, implies that all dynamical
features in the replicator model depend on the fitness matrix $A$
through differences between its elements. As a matter of fact, we find that
a set of relevant quantities in the characterization of such
features is given by the differences $g_{ij}=a_{ij}-a_{jj}$. They
compare the contribution of species $j$ to the evolution of $i$
with its self-contribution. We call $g_{ij}$ the {\it attitude} of
species $j$ toward species $i$. A positive attitude, $g_{ij}>0$,
indicates that species $j$  contributes to the growth of $i$ more
than to its own growth. It is therefore an {\it altruistic}
attitude. On the other hand, $g_{ij}<0$ stands for an {\it
egoistic} attitude. By  definition,  the ``self-attitudes''
$g_{jj}$ are all zero. As we show in the following, the stability
properties of many of the equilibria of Eqs.~(\ref{sys}) can be
given in terms of the attitudes $g_{ij}$.

For large $N$, Eqs.~(\ref{sys}) have generally a large number of
fixed points. They can be classified by their number of
non-vanishing coordinates, i.e. by the number of surviving
species at each fixed point. There are $N$ fixed points with only
one surviving species. The coordinates of these points are all
zero except that corresponding to the surviving species, which
equals unity. We  emphasize that such equilibrium points,
situated at the {\it vertices} of $S_N$,  always exist,
independently of the values of the coefficients $a_{ij}$. In
contrast, the number of  fixed points with $n$ surviving species
($1<n<N$) depends on $a_{ij}$. We do know, however, that
$C(N,n)=N!/n!(N-n)!$ is an upper bound for this number. These
equilibrium points lie on the {\it edges} of the polyhedron
$S_N$. Finally, only one equilibrium with $N$ surviving species
can exist \cite{hof}. We call it the {\it coexistence} fixed
point. If it exists, its coordinates  satisfy $\sum_j a_{ij}
x_j=\sum_{kl} a_{kl} x_k x_l$ for all $i$.

For a general fitness matrix $A$ and arbitrary $N$, the linear
stability analysis of the fixed points of the replicator
equations is quite a difficult task. Still, it is possible to
find the general stability conditions for vertices. The $j$-th
vertex, with $x_j=1$ and $x_i=0$ for all $i\neq j$, is linearly
stable if and only if $g_{ij}<0$ for all $i\neq j$, i.e. if
species $j$ is egoistic toward all other species. Note that this
condition can be simultaneously verified for several --or even
all-- species, in which case the system is multistable.

An important case where the stability analysis can be fully
achieved for all the fixed points is that of a diagonal matrix
($a_{ij}=0$ for $i\neq j$), which will be used as a reference
case in our statistical analysis of survival and extinction. In
this case, as above, the $j$-th vertex is linearly stable if and
only if $g_{ij}<0$ or, equivalently, $a_{jj}>0$. The equilibria
on the edges of $S_N$, on the other hand, are always unstable.
Meanwhile, the coexistence point is stable if and only if
$g_{ij}=-a_{jj}>0$  for all $i$, i.e. if all species are
altruistic toward all other species. Consequently, if the
coexistence point is stable, it is the only stable equilibrium,
since the vertices are in turn unstable. The global stability of
the coexistence point follows also from a general theorem for the
replicator system with symmetric fitness matrices \cite{hof}.

\begin{figure}[h]
\centerline{\psfig{file=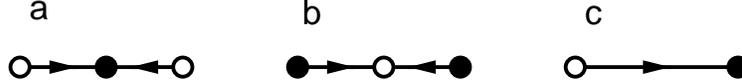,width=10cm,angle=0,clip=}}
\caption{Dynamical behaviors for a two-species system: (a) coexistence,
(b) bistability, and  (c) domination by one species. In the three cases,
empty dots indicate unstable fixed points, while full dots indicate
stable equilibria.} \label{f1}
\end{figure}

Linear stability analysis can also be carried out explicitly for
small values of $N$. For two species, the dynamics is effectively
one-dimensional. The phase space $S_2$ can be mapped onto the
interval $[0,1]$, where one of the variables (say $x_1$) takes on
its values, while $x_2=1-x_1$ at all times. In Fig.~\ref{f1}  we
sketch the three possible dynamical scenarios. Coexistence
(Fig.~\ref{f1}a) occurs when both species have positive attitude
toward each other, while bistability (Fig.~\ref{f1}b) occurs when
both species have  negative attitudes. Domination by one of the
species takes place when one species is altruistic and the other
species is egoistic. In such case the egoistic species dominates,
while the other becomes extinct. For three species the dynamics
is effectively bidimensional, and there is a wider spectrum of
dynamical behaviors, including the possibility of existence of
heteroclinic cycles \cite{hof}. In general, we find that
stability properties depend not only on the signs of the
attitudes $g_{ij}$, but also on other combinations of the
elements $a_{ij}$.

Real ecosystems, however, are known to involve at least several
species with complex patterns of interactions \cite{eco1,eco2}.
The restricting conditions that make the replicator model
analytically solvable fall well apart from any situation with
biological significance. We therefore turn our attention to
many-species systems with nontrivial interactions. In order to
explore an ample class of fitness matrices, we choose the
elements $a_{ij}$ at random, from prescribed distributions. This
choice calls for a statistical analysis of the results, which in
turn require resorting to the numerical resolution of the
replicator equations. In the next section we discuss methods and
present results of such analysis.

\section{Random fitness matrices for many-species ecosystems}

As advanced above, we analyze in this section the dynamics of the
replicator model with a random fitness matrix and for a large
number of species. In a generic situation, the coexistence
equilibrium will be unstable and a substantial number of species
will become extinct. We are particularly interested in studying
the dependence of the number of surviving species on the
distribution  of the elements $a_{ij}$. Our ultimate goal would be
to identify the ecological attitude that a species should have in
order to optimize its probability of surviving within a given
population, i.~e. for a given distribution  of $a_{ij}$. From
then on, in order to make the description of interactions more
compact, we characterize a species $j$ by its mean attitude
\begin{equation}
G_j= \frac{1}{N}\sum_i g_{ij}=\frac{1}{N}\sum_i (a_{ij}-a_{jj}).
\end{equation}
Depending on the distribution of $a_{ij}$, the mean attitudes in
a given population will take values in different ranges. For each
realization of the evolution of the system, we respectively call
$n(G)$ and $n_s(G)$ the (normalized) distribution of the number of
species and of the number of surviving species on $G$. Namely,
for small $\delta G$, the products $N n(G)\delta G$ and $Nn_s(G)
\delta G$ represent, respectively, the total number of species and
the number of surviving species in the interval $(G,G+\delta G)$.
The average fraction of surviving species as a function of the
mean attitude is given by
\begin{equation}
\rho(G)=\frac{\langle n_s(G) \rangle}{\langle n(G) \rangle},
\end{equation}
where $\langle \cdot \rangle$ indicates average over realizations,
with different $a_{ij}$ and different initial conditions. The
product $\rho(G)\delta G$ measures the survival probability of a
species with mean attitude between $G$  and $G+\delta G$ within
the given population.

Taking into account the symmetries of the replicator equations
(\ref{sys}), discussed in Sect. \ref{ii}, we have selected a set
of distributions for $a_{ij}$ that results to be representative
and interesting. In most cases, we fix a significant subset of
the matrix elements equal to zero, while the remaining $a_{ij}$
are uniformly distributed on a finite interval. The subset of
null elements can be fixed deterministically or at random over
the matrix, and defines a reference level for all the other
$a_{ij}$. Three intervals for the uniform distribution of the
nonzero elements will be considered, namely,
$(-\frac{3}{2},-\frac{1}{2})$, $(-\frac{1}{2},\frac{1}{2})$, and
$(\frac{1}{2},\frac{3}{2})$. In this way, we take into account
translations of the distribution of the nonzero elements with
respect to the reference value. Note that, on the other hand, the
three intervals have the same width. We briefly discuss the
effect of changing this width at the end of Sect.~\ref{4.2}.

A special reference case which will also be considered in the
following is a fitness matrix where no element is {\it a priori}
fixed to zero, and $a_{ij}$ is chosen from the uniform
distribution for all $i$ and $j$. This special case is
statistically invariant under translations of the interval and
changes of its width --up to a modification in the time scale.

\subsection{Numerical calculations}

In the numerical resolution of the replicator equations we face a
main difficulty, namely, the dynamical instability of the
invariant set $S_N$. From Eqs.~(\ref{uno}) it follows that,
depending on the sign of the global coupling term $\sum_j x_j
f_j=\sum_{j,k} a_{jk} x_jx_k$, a small perturbation on the
condition $\sum_i x_i=1$ may be amplified by the dynamics, driving
the system away from $S_N$. This is expected to happen in the
numerical calculations, due to rounding-off and discretization
effects. Using a second-order Runge-Kutta (RK) integration
algorithm \cite{NR}, we have  observed departures from $S_N$ for
very long times only, typically when the evolution is very slow
and the system is close to an unstable fixed point. However, such
events prompt us to modify the integration algorithm, in order to
ensure permanence on $S_N$ at all times. To solve the problem, we
have included a normalization operation at each time step. After
computing the updated variables  $x_i(t+dt)$ using the RK
algorithm, we compute the sum $\sigma=\sum_i x_i(t+dt)$ and
renormalize the variables as $x_i(t+dt) \to x_i(t+dt)/\sigma$. We
have exhaustively tested this algorithm and observed that, in the
regions where the original RK algorithm does not produce
divergences, the solutions by both methods are indistinguishable.
Furthermore, for the original RK algorithm we have observed that
the perturbations that drive the solution away from $S_N$ are
initially of the order of the working precision ($\sim
10^{-16}$), and the corrections introduced by the normalization
procedure are of the same order, or $\sim10^{-15}$ at most. Such
modifications are therefore much smaller than the discretization
error of the RK algorithm, and the effect of normalization on the
numerical precision results to be negligible.

We have chosen initial conditions for $x_i$ with uniform
distribution over $S_N$. They have been generated by drawing
$N-1$ uniformly distributed random numbers $y_i$ in $(0,1)$, with
$y_1<y_2< \cdots<y_{N-1}$, and defining $x_i(0)=y_{i+1}-y_i$ for
$1\le i<N$ and $x_N(0)=1-y_{N-1}$.

After integration over sufficiently long times, the values of
$x_i$ corresponding to the species which are in the way of
becoming extinct have typically reached extremely low levels.
Such levels are usually  meaningless from the  biological
viewpoint. In  fact, the discreteness of biological populations
fixes a lower bound of  $( \sum_j n_j )^{-1}$ for each $x_i$
\cite{AZ}. Since the replicator model does not keep trace of the
total population in the ecosystem, we choose as a threshold for
extinctions $x_{\min}=10^{-3}/N$. If, at the end of the
integration, a variable $x_i$ has reached a value below
$x_{\min}$, species $i$ is considered to have undergone
extinction. To study the evolution of the number of surviving
species, this definition is extended to all times. Surviving
species at time $t$ are hence defined as those for which
$x_i(t)>x_{\min}$. Our value of $x_{\min}$ is essentially
arbitrary, but we find that the results are robust with respect
to variations in this extinction threshold.

Our calculations correspond mostly to the case $N=32$, though we
have verified that our main qualitative conclusions are valid for
considerably larger ecosystems.  As indicated above, our
statistical analysis requires to average results over
realizations with different initial conditions and different
matrix elements. Averages were taken over sets of $900$ to  $6000$
realizations. In the following, we present the main results of our
analysis, which focuses on the evolution of the number of
surviving species and on the determination of the survival
probability as a function of the mean attitude of each species.

\subsection{Diagonal and full-matrix systems}

We begin by analyzing, as a reference situation, the case of a
diagonal fitness matrix ($a_{ij}=0$ for $i\neq j$). Recall that
the stability properties in this case can be worked out
exhaustively (Sect. \ref{iii}). Note also that, for the diagonal
system, $g_{ij}=-a_{jj}$ is independent of $i$, such that the
attitude of each species $j$ toward the remaining species is
uniform.

\begin{figure}[h]
\centerline{\psfig{file=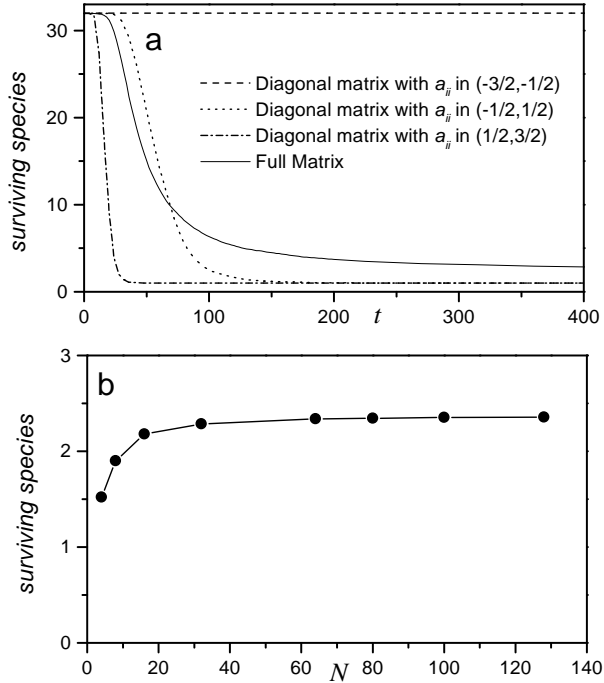,width=8cm,angle=0,clip=}}
\caption{(a) Number of surviving species as a function of time for diagonal
fitness matrices and for the full-matrix case, with nonzero
elements uniformly distributed on the indicated intervals. (b)
Long time limit of the number of surviving species in the
full-matrix case as a function of $N$.}
\label{f2}
\end{figure}

When the diagonal elements are drawn at  random from then interval
$(-\frac{3}{2},-\frac{1}{2})$, all these elements are negative.
In this case all  the species are altruistic, the coexistence
point is globally stable, and all the species survive. The
survival probability $\rho (G)$ is thus a flat distribution. In
the case that the diagonal elements are drawn from $(-\frac{1}{2},
\frac{1}{2})$, typically half of the  species --those with
$a_{jj}<0$-- are altruistic toward any other species, while the
other half are egoistic. The only stable equilibria are the
approximately $N/2$ vertices corresponding to the survival of a
single species with $a_{jj}>0$. The surviving species is selected,
in each realization, by the initial condition. Finally, when the
diagonal elements are chosen from  $(\frac{1}{2},\frac{3}{2})$,
all the species are egoistic and the $N$ vertices  are stable.
These are again the only stable equilibria and only one species
survives. Figure \ref{f2}a shows the evolution of the average
number of surviving species for the above three cases. The
corresponding survival probability distributions $\rho (G)$ will
be discussed later on, for comparison with other random matrices.

A second case of reference is given, as advanced above, by the
situation where no elements of the fitness matrix are {\it a
priori} fixed to zero. We refer to this situation as the
full-matrix case. The evolution of the average number of
surviving species for the full-matrix case is also shown in
Fig.~\ref{f2}a. It is interesting to point out the relatively low
number of species that ultimately survive in  this case, which
for $32$ species is around $2.28$ on the average. In Fig.
\ref{f2}b we show the average number of surviving species for the
full-matrix system as a function of $N$. It does not change
significantly with $N$, since for $N=128$ it is around $2.35$. As
for the diagonal matrix, the distribution $\rho(G)$ for the
full-matrix case will be presented later.

\subsection{Systems with multi-diagonal matrices} \label{4.2}

Taking the diagonal-matrix and the full-matrix systems as limiting
reference cases, we now  study  intermediate, more complex
situations. We consider Eqs.~(\ref{sys}) for matrices with
different numbers of nonzero diagonals. We define a $k$-diagonal
random matrix, with $k$ an odd natural number, as a matrix whose
elements $a_{ij}$ vanish for $|i-j|> (k-1)/2$ and which are
otherwise taken at random from a given distribution. Clearly,
$k=1$ corresponds  to the diagonal matrix, while the full-matrix
case corresponds to $k=2N-1$.

\begin{figure}[h]
\centerline{\psfig{file=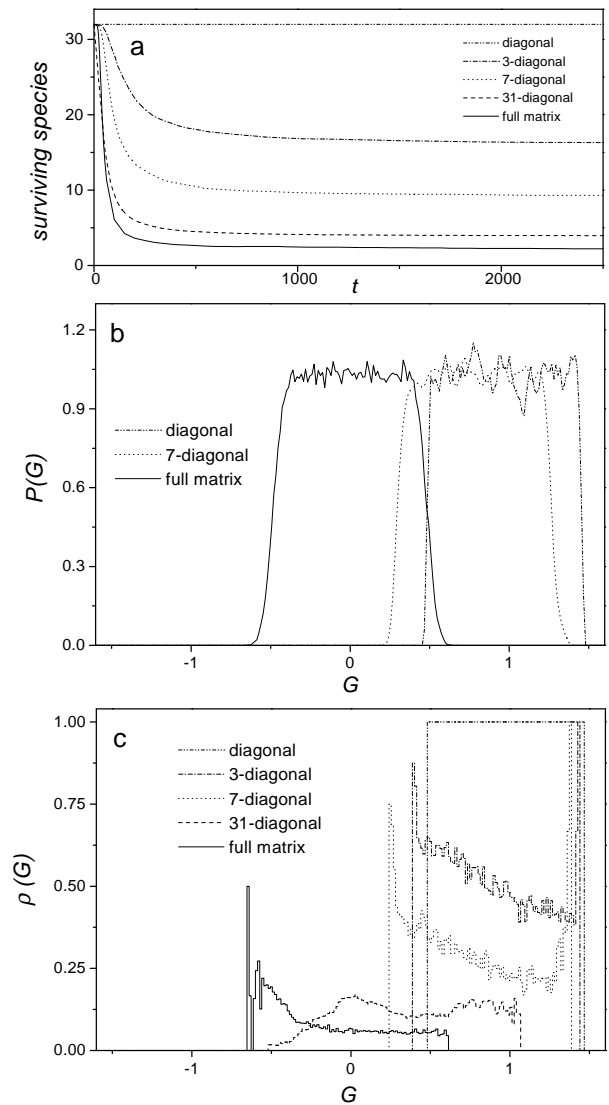,width=8cm,angle=0,clip=}}
\caption{(a) Number of surviving species as a function of time, (b)
distributions of mean attitudes $P(G)$, and (c) survival
probability $\rho(G)$, for multi-diagonal fitness matrices with
nonzero elements taken from a homogeneous distribution on
$(-\frac{3}{2},-\frac{1}{2})$.}
\label{f3}
\end{figure}

Figure \ref{f3}a  shows the evolution of the mean number of
surviving species for multi-diagonal matrix systems with several
values of $k$, and with the nonzero elements distributed
uniformly on the interval  $(-\frac{3}{2},-\frac{1}{2})$. It can
be seen that the average number of extinctions rapidly increases
with the number of nonzero diagonals. In Fig.~\ref{f3}b we show
the average normalized distribution of mean attitudes $\langle
n(G) \rangle$ for three of the five cases analyzed in
Fig.~\ref{f3}a. The range of mean attitudes where $\langle n(G)
\rangle$ is sensibly different from zero moves toward the left as
$k$ is increased, reflecting the fact that the number of egoistic
species grows. Combining this observation with the results of
Fig.~\ref{f3}a we conclude that the more egoistic the population
is, the less the number of surviving species.

In Fig.~\ref{f3}c we show the numerical result for $\rho(G)$ in
the same realizations of Fig.~\ref{f3}a. Since in the
diagonal-matrix case ($k=1$) all the species survive, $\rho(G)$
corresponds to a uniform distribution. The results for $k=3$ and
$k=7$ are remarkably different. The survival probabilities  are
in general lower than in the diagonal-matrix case and decrease
almost linearly with the mean attitude over a wide range of
values of $G$. Near the ends of the relevant intervals the linear
behavior breaks down, and $\rho(G)$ shows abrupt maxima. The
maximum on the left side of the intervals indicate that, as
expected, being one of the most egoistic species is a good
strategy, as far as survival is concerned. Less obviously, the
rightmost maximum indicates that being one of the most altruistic
species is also convenient for survival. For $k=15$, $\rho(G)$
does not attains a maximum on the left extreme  of the relevant
interval and an intermediate optimum value of the mean attitude,
close to $G=0$ appears, instead.

\begin{figure}[h]
\centerline{\psfig{file=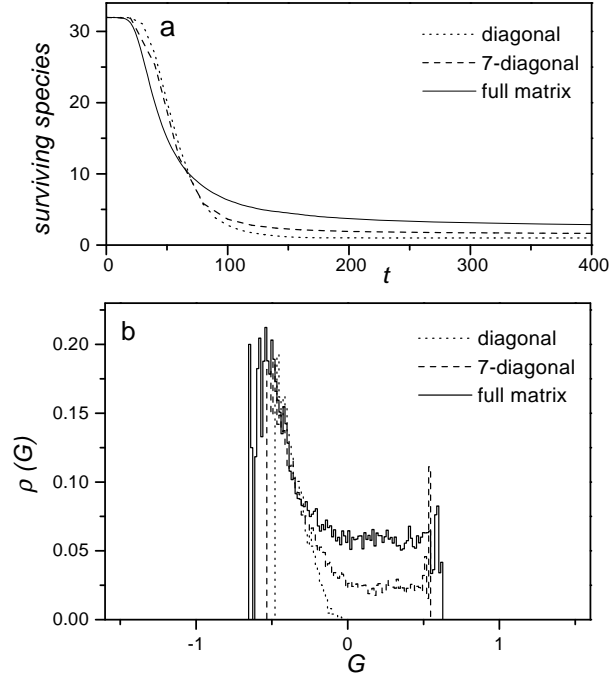,width=8cm,angle=0,clip=}}
\caption{(a) Number of surviving species as function of time and (b)
survival probability $\rho(G)$  for random multi-diagonal
matrices, with nonzero elements uniformly distributed on
$(-\frac{1}{2},\frac{1}{2})$.} \label{f4}
\end{figure}

\begin{figure}[h]
\centerline{\psfig{file=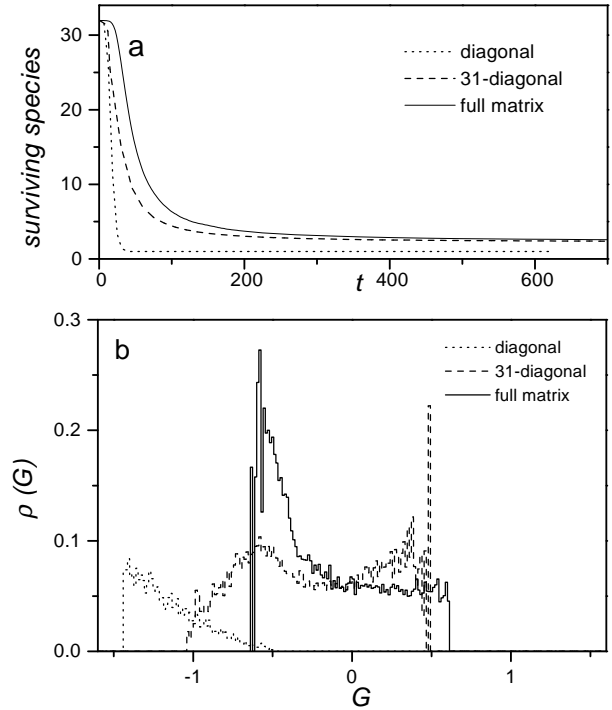,width=8cm,angle=0,clip=}}
\caption{(a) Number of surviving species as a function of time and
(b) survival probability $\rho(G)$,  for random multi-diagonal
matrices with the nonzero elements uniformly distributed on
$(\frac{1}{2},\frac{3}{2})$. Note that, for $31$ diagonals, more
than $1/4$ of the matrix elements are nonzero. Even though, the
number of surviving species is nearly the same as in the full
matrix case, which corresponds to $k=2N-1=63$.}
\label{f5}
\end{figure}

In Figs.~\ref{f4} and \ref{f5} we present the results for
multi-diagonal matrices with their non-vanishing elements
distributed uniformly on $(-\frac{1}{2},\frac{1}{2})$ and
$(\frac{1}{2},\frac{3}{2})$, respectively. In these situations,
the population is more egoistic than in the case of Fig.~\ref{f3},
and the number of surviving species is very small. In contrast
with that case, the number of survivals in the diagonal-matrix
system is lower than for the full-matrix system. Now, only one
species survives for a diagonal matrix, and the full-matrix case
appears to be the situation with the highest rate of survivals. In
the case of Fig.~\ref{f4}, the transition between the limiting
situations is smooth: the profiles of $\rho(G)$ for different
values of $k$ are similar. All of them decrease monotonically
with the mean attitude for intermediate values of $G$, and
present abrupt maxima at the ends of the relevant intervals. The
situation in Fig.~\ref{f5} is more complex, and somehow more
similar to the case of Fig.~\ref{f3}, since a maximum of $\rho$
for an intermediate value of $G$ is observed for relatively high
values of $k$.

\begin{figure}[h]
\centerline{\psfig{file=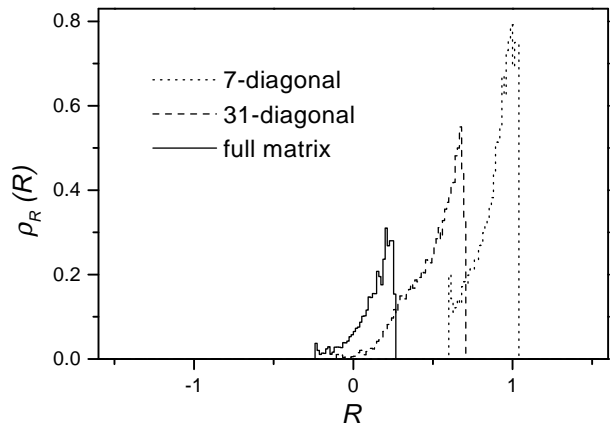,width=8cm,angle=0,clip=}}
\caption{Fraction of surviving species $\rho_R$ as function of
$R$ for a $7$-diagonal matrix with nonzero elements uniformly
distributed on$(-\frac{1}{2},\frac{1}{2})$.} \label{f6}
\end{figure}

As a function of $G$, the fraction of survivals $\rho$
characterizes the probability of surviving for a given species in
terms of its ecological attitude toward other species in the
system. It is also possible to characterize the survival
probability in terms of the average attitude of the ecosystem
toward a given species $j$, which we define as
\begin{equation}
R_j = \frac{1}{N}\sum_i (a_{ji}-a_{ii})=\frac{1}{N}\sum_i g_{ji}.
\end{equation}
Thus, we have also studied $\rho_R (R)$, the average fraction of
surviving species for a given value of $R$. Figure \ref{f6}
presents the corresponding results for $k$-diagonal systems with
nonzero elements uniformly distributed on
$(-\frac{3}{2},-\frac{1}{2})$ and different values of $k$. We
observe a monotonous growth of $\rho_R$, clearly indicating that
the probability of surviving for a given species is higher when
the remaining species are more altruistic toward it.

\begin{figure}[h]
\centerline{\psfig{file=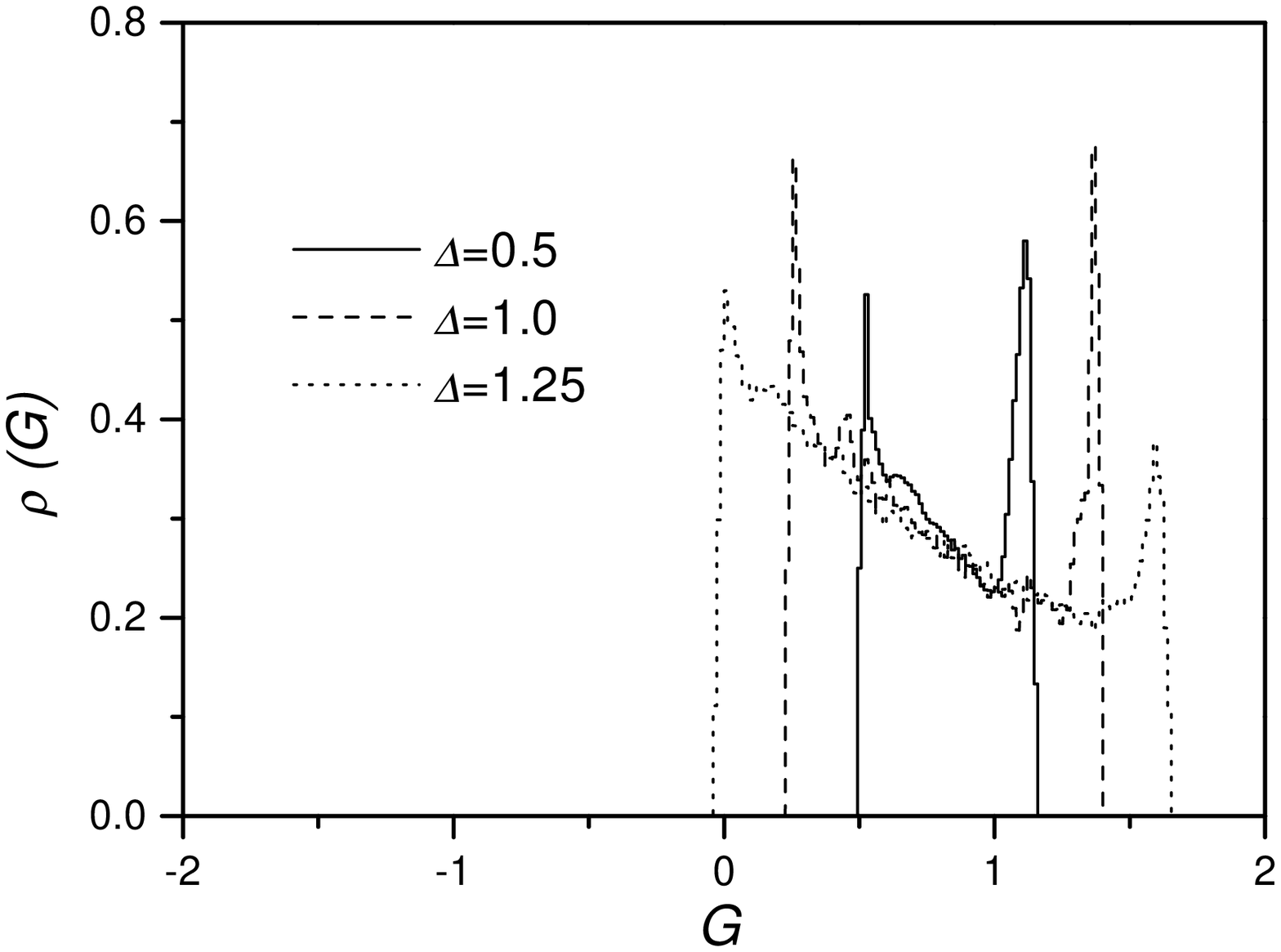,width=8cm,angle=0,clip=}}
\caption{Survival probability $\rho(G)$ corresponding to $7$-diagonal matrices
with nonzero elements uniformly distributed on intervals of
different widths $\Delta$ around $-1$.}\label{f10}
\end{figure}

Up to now, we have kept the width of the distribution of nonzero
elements equal to unity. Here, we briefly analyze the effect of
changing this parameter in a particular situation. We consider a
$7$-diagonal matrix with its nonzero elements distributed
uniformly in an interval of width $\Delta$ around $-1$. Numerical
calculations show that the average number of extinctions as a
function of time is practically independent of $\Delta$. Figure
\ref{f10} shows $\rho(G)$ for different values of $\Delta$. As
expected, the relevant interval of attitudes grows as the width
of the distribution increases. However, no qualitative changes
are observed in the profile of $\rho$.

\subsection{Disordered random matrices}

Multi-diagonal fitness matrices represent interaction patterns for
which the ensemble of species can be ordered in a well-defined
sequence, where each species interacts with a few neighbors
--besides the global interaction that couples the whole
ecosystem. We now turn the attention to systems where the
interaction pattern is more disordered than in the multi-diagonal
case. We first consider fitness matrices where the diagonal
elements are uniformly distributed in a certain interval, while
the non-diagonal elements are drawn from the same distribution
with probability $p$ and fixed to zero with probability $1-p$.
The diagonal-matrix case corresponds to $p=0$ and the full-matrix
system is attained for $p=1$.

\begin{figure}[h]
\centerline{\psfig{file=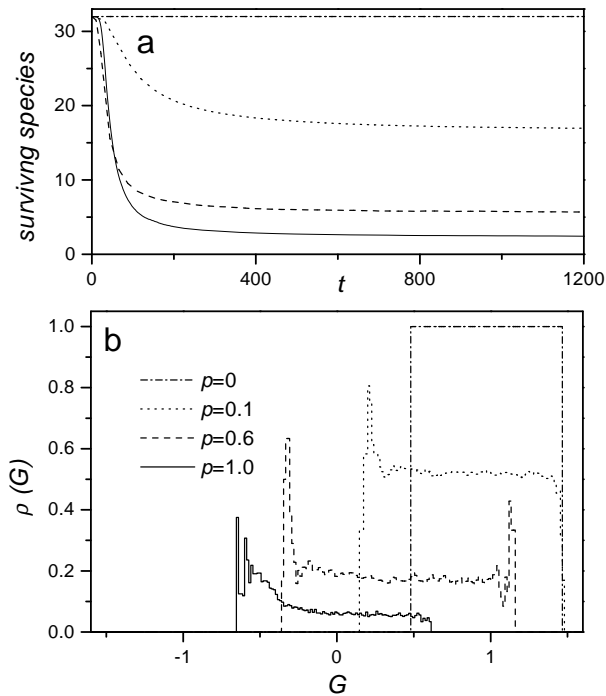,width=8cm,angle=0,clip=}}
\caption{(a) Number of surviving species as a function of time
and  (b)  survival probability $\rho(G)$, for random matrices with
diagonal elements taken from homogeneous distributions on
$(-\frac{3}{2},-\frac{1}{2})$, and non-diagonal matrix elements
taken from the same distribution with probability $p$ and fixed
as zero with probability $1-p$. Note that $p=0$ corresponds to the
diagonal matrix case while $p=1$ corresponds to the full matrix
case.}  \label{f7}
\end{figure}

In Fig.~\ref{f7} we show results for the average number of
surviving species as a function of time, and for the distribution
$\rho(G)$, in a system where the nonzero matrix elements are
taken from  a uniform distribution on $(-\frac{3}{2},-\frac{1}{2})$,
and for different values of $p$. We find that the distributions
$\rho(G)$ for $0<p<1$ regularly interpolate the two extreme
cases. The same feature is found when the nonzero
elements are taken from $(-\frac{1}{2},\frac{1}{2})$ and
$(\frac{1}{2},\frac{3}{2})$.

\begin{figure}[h]
\centerline{\psfig{file=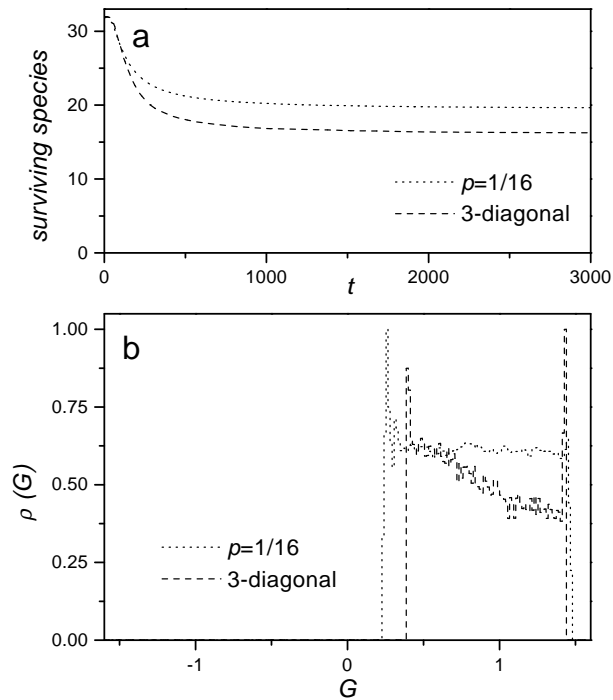,width=8cm,angle=0,clip=}}
\caption{(a) Number of surviving species as a function of time and (b)
survival probability $\rho(G)$,  for a 3-diagonal matrix whith
nonzero elements uniformly distributed on
$(-\frac{3}{2},-\frac{1}{2})$ and for a random disordered matrix
with diagonal elements taken from the same distribution, and
non-diagonal elements taken with probability $p$ from the same
distribution and fixed as zero with probability $1-p$.}
\label{f8}
\end{figure}

In Fig.~\ref{f8} we compare the results for a $3$-diagonal system
and for a disordered system  with  probability $p=2/N=1/16$,
which correspond to the same average number of nonzero matrix
elements. It can be seen that the number of surviving species is
significantly higher in the disordered system. Note that both
systems have the same distribution of attitudes $\langle n(G)
\rangle$. This is because, by definition, this function does not
depend on the ordering of the interactions, but only on the
average number, and distributions, of the nonzero matrix
elements. Hence, the difference on the number of extinctions from
both systems is a pure consequence of the disorder of the
interactions.

Finally, we study the replicator for a particular set of matrices
that constitutes a continuous interpolation between multi-diagonal
and disordered matrices. We consider a $k$-diagonal fitness matrix
and, with probability $q$, we set to zero each element of the
$k-1$ nonzero sub-diagonals. For each of these vanishing elements,
we add a new random element outside the multi-diagonal block,
$|i-j| >(k-1)/2$. In this way it is possible, for instance, to
transit continuously between the two cases reported in
Fig.~\ref{f8}, starting from a $3$-diagonal matrix ($q=0$) and
increasing the probability $q$ to approach the disordered case
($q=1$). Note that the distribution $\langle n(G) \rangle$ does
not depend on $q$, and the only difference between systems with
different values of this parameter is the degree of disorder in
their interactions. In Fig.~\ref{f9} we present results for this
construction, but starting from a $7$-diagonal matrix system. It
can be seen that, as the degree of disorder in the fitness matrix
grows, the number of survivals increases and the distributions
$\rho(G)$ become flatter.

\section{Conclusions}

In this paper we have presented a statistical study of survival
and extinction in the replicator model with linear fitness,
exploring in particular the effect of the pattern of interactions
in the ecosystem. We have selected a set of interaction patterns
and  fitness matrices that is expected to provide a representative
(though not complete) sample of the different possible situations
found in complex model ecosystems. Through the definition of the
mean attitude of a species, we give a compact   characterization
of its ecological behavior, avoiding the problem of dealing with
the detailed information involved in the fitness matrix. Our
analysis has been focused on the study of the evolution of the
number of surviving species for different interaction patterns,
and on the determination of the survival probability as a
function of the mean attitude. In the following, we summarize and
comment our main conclusions.

\begin{figure}[h]
\centerline{\psfig{file=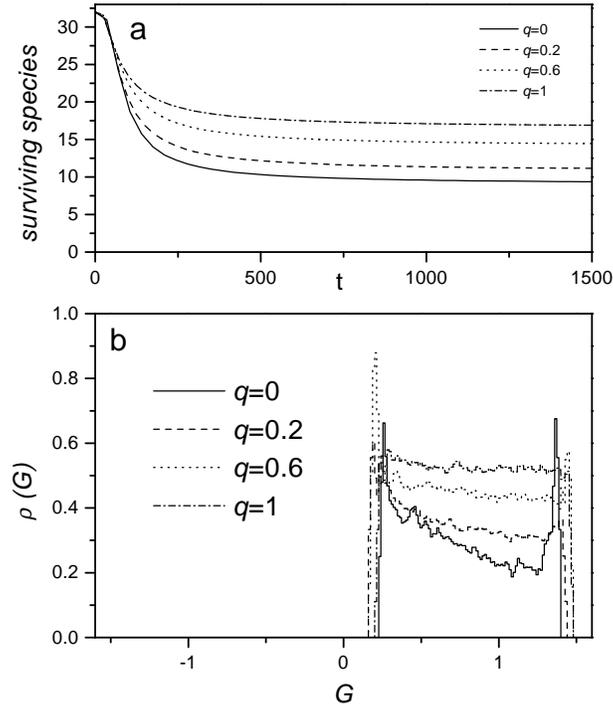,width=8cm,angle=0,clip=}}
\caption{(a) Number of surviving species as a function of time and (b) survival
probability $\rho(G)$,  for a 7-diagonal system that is being
disordered by increasing the probability $q$. The nonzero matrix
elements are uniformly distributed on
$(-\frac{3}{2},-\frac{1}{2})$.} \label{f9}
\end{figure}

Concerning the number of surviving species in a given population,
we have found that, in all cases, it increases as the mean
attitude of the whole ensemble of species shifts to larger values.
In other words, survival probabilities are higher in altruistic
populations. This property is clearly observed for multi-diagonal
and disordered matrices, and for different distributions of the
nonzero matrix elements.

The survival probability grows also with the disorder of the
fitness matrix, as illustrated by Figs.~\ref{f9} and \ref{f10},
where we have compared the results for populations with different
degrees of disorder in their interactions, but equal
distributions of the mean attitude. In contrast with the case of
multi-diagonal fitness matrices, where the number of interaction
links is the same for all species, for disordered matrices there
is a chance that the ecosystem becomes divided into weakly
interacting subsystems. Within one of such subsystems, each
species interacts with a relatively smaller population and its
survival probability thus increases.

Once the fitness matrix of a population is fixed, we find that
the survival probability is higher for the most extreme attitudes.
In most situations, in fact, species with the lowest and largest
mean attitudes --i. e., the most egoistic and the most altruistic,
respectively-- have better chances of surviving. Some remarkable
exceptions have been found, however, where the egoistic extreme is
not a good option, and an intermediate mean attitude maximizes the
survival probability (see Fig.~\ref{f5}). In contrast, the
altruistic extreme has always been found to have a relative high
chance, if not the highest, of surviving.

Let us finally stress that the present analysis has explored only
a class of possible interaction patterns in the multi-species
replicator model, with specific choices for the structure of the
fitness matrix and the distribution of its elements. The
possibility is open for the consideration of much more general
situations, approaching the statistical description of large real
ecosystems with arbitrary, even highly complex, food-web
structures.

\section*{Acknowledgement}

This work has been partially carried out at the Abdus Salam
International Centre for Theoretical Physics (Trieste, Italy).
The authors thanks the Centre for hospitality.

\end{document}